\documentclass[
reprint,
superscriptaddress,
showpacs,showkeys,floatfix,
amsmath,amssymb,
aps,
jcp
]{revtex4-1}

\usepackage{graphicx}
\usepackage[ colorlinks,
    linkcolor={blue},
    citecolor={blue},
    urlcolor={blue}]{hyperref}

\usepackage{amsmath}
\usepackage{amssymb}
\usepackage{dcolumn}
\usepackage{bm}

\usepackage[english]{babel}
\usepackage{microtype}
\usepackage{indentfirst}
\usepackage{siunitx, booktabs}
\usepackage{tabularx}
\usepackage[caption=false]{subfig}
\usepackage{array}
\usepackage{setspace}
\usepackage{braket}
\usepackage{xcolor}
\usepackage{placeins}

\usepackage{amsfonts}
\usepackage{algorithmic}
\usepackage{textcomp}
\usepackage{xcolor}
\usepackage{xspace}
\usepackage{url}
\usepackage{listings}

\makeatletter
\newcommand\xleftrightarrow[2][]{%
  \ext@arrow 9999{\longleftrightarrowfill@}{#1}{#2}}
\newcommand\longleftrightarrowfill@{%
  \arrowfill@\leftarrow\relbar\rightarrow}
\makeatother

\begin{document}

\title{Water structure near the surface of Weyl semimetals as catalysts in photocatalytic proton reduction}



\author{Jure Gujt}
\affiliation{Dynamics of Condensed Matter and Center for Sustainable Systems Design, Chair of Theoretical Chemistry, Paderborn University, Warburger Str. 100, D-33098 Paderborn, Germany}%
\author{Peter Zimmer}
\affiliation{Chair for Inorganic Chemistry of Sustainable Processes and Center for Sustainable Systems Design, Department of Chemistry, Paderborn University, Warburger Str. 100, D-33098 Paderborn, Germany}
\author{Frederik Zysk}
\affiliation{Dynamics of Condensed Matter and Center for Sustainable Systems Design, Chair of Theoretical Chemistry, Paderborn University, Warburger Str. 100, D-33098 Paderborn, Germany}%
\author{Vicky S\"u{\ss}}
\affiliation{Max Planck Institute for Chemical Physics of Solids, N\"othnitzer Str. 40, D-01187 Dresden, Germany}%
\author{Claudia Felser}
\affiliation{Max Planck Institute for Chemical Physics of Solids, N\"othnitzer Str. 40, D-01187 Dresden, Germany}%
\author{Matthias Bauer}
\email{matthias.bauer@upb.de}
\affiliation{Chair for Inorganic Chemistry of Sustainable Processes and Center for Sustainable Systems Design, Department of Chemistry, Paderborn University, Warburger Str. 100, D-33098 Paderborn, Germany}%
\author{Thomas D. K\"uhne}
\email{tdkuehne@mail.upb.de}
\affiliation{Dynamics of Condensed Matter and Center for Sustainable Systems Design, Chair of Theoretical Chemistry, Paderborn University, Warburger Str. 100, D-33098 Paderborn, Germany}%

\date{\today}

\verb

\begin{abstract}
In this work, second-generation Car-Parrinello-based QM/MM molecular dynamics simulations of small nanoparticles of NbP, NbAs, TaAs and 1T-TaS$_2$ in water are presented. The first three materials are topological Weyl semimetals, which were recently discovered to be active catalysts in photocatalytic water splitting. The aim of this research was to correlate  potential differences in the water structure in the vicinity of the nanoparticle surface with the photocatalytic activity of these materials in light induced proton reduction. The results presented herein allow to explain the catalytic activity of these Weyl semimetals: the most active material, NbP, exhibits a particularly low water coordination near the surface of the nanoparticle, whereas for 1T-TaS$_2$, with the lowest catalytic activity, the water structure at the surface is most ordered. In addition, the photocatalytic activity of several organic and metalorganic photosensitizers in the hydrogen evolution reaction was experimentally investigated with NbP as proton reduction catalyst. Unexpectedly, the charge of the photosensitizer plays a decisive role for the photocatalytic performance. 
\end{abstract}

\maketitle 

%


%
%

%

\section{Introduction}
\label{introduction}

Catalysis nowadays plays a very important role in almost every field of chemistry. Owing to its complexity, the understanding of the fundamental processes itself is nevertheless still a major challenge~\cite{fpaper01,fpaper02,fpaper03,fpaper04}. An exemplary catalytic process with increasing importance is proton reduction, which employs solar energy to produce molecular hydrogen (H$_2$) with a high potential in terms of "green energy"~\cite{fpaper05,fpaper06, mb1}. In this redox process, a metal or a semiconductor is usually employed as a catalyst, which demands a stable supply of itinerant electrons be delivered to the surface. A dye or photosensitizer, for instance, can deliver such high-energy electrons after excitation with light. To understand this type of catalysis, the interaction of the photosensitizers, as well as that of all educts, products and solvents with the surface of the catalysts is crucial, especially when a heterogeneous catalyst is involved \cite{fpaper07,fpaper08,fpaper09}. In addition, for a semiconducting catalyst, high mobility of electrons and holes is desired in order to reduce the probability of recombination of electron-hole pairs that are created during the redox process. Rajamathi et al. have investigated Weyl topological semimetals as catalysts for the catalytic hydrogen evolution reaction (HER)~\cite{fpaper}. A fundamental property of Weyl and Dirac semimetals is their high carrier mobility, which arises from the linear bands of the Dirac cone ~\cite{fpaper10,fpaper11}. Furthermore, thanks to their robust and topologically protected surface states, these materials avoid surface contamination, which is the bottleneck in such catalytic transformations~\cite{fpaper12,fpaper13}. An essential property of a topological insulator or Weyl semimetal is an energy band inversion, which is known in chemistry as inert pair effect~\cite{fpaper14}. The inert pair effect can be observed in many compounds containing heavy metals. Since band crossing is forbidden in relativistic band structures, in topological insulators a new bandgap opens. As a result, a surface state having a Dirac cone in the electronic structure appears. Dirac and Weyl semimetals form the transition between topological and trivial insulators. In a Weyl semimetal, pairs of Dirac cones are formed in the bulk of the material, whereby the number of pairs depends on the detailed symmetry of the particular semimetal~\cite{PhysRevB.92.115428}.
The present work expands on the work by Rajamathi et al.~\cite{fpaper}, where the HER activity of various transition-metal monopnictides as proton reduction catalysts was investigated. Therein, a decreasing activity in the order NbP $>$ TaP $>$ TaAs $>$ NbAs was found. Also, the Gibbs free energy of the hydrogen absorption $\Delta$G$_{H^*}$ was calculated and related to the volcano plot (Fig.~2c in Ref.~\cite{fpaper}) that revealed a higher catalytic activity with $\Delta$G$_{H^*}$ being closer to zero. 

To explain the enhanced catalytic reactivity, in this work, we investigate the impact of the water structure around these nanoparticles utilizing second-generation Car-Parrinello-based QM/MM molecular dynamics simulations in aqueous solution. Specifically, we study three transition-metal monopnictides NbP, NbAs and TaAs with negative $\Delta$G$_{H^*}$ and 1T-TaS$_2$ with positive $\Delta$G$_{H^*}$, respectively. Furthermore, the effect of different organic and metalorganic dyes \cite{dye1, dye2, dye3} on the photocatalytic proton reduction activity of NbP, which is the most active monopnictide considered here, is analyzed experimentally.

\section{Computational details}
\label{compdet}

All simulations were performed with the cp2k software package~\cite{cp2kpaper, CP2Krev}. The system consisted of a single nanoparticle and water molecules in a periodic cubic simulation box with an edge length of 32~\AA. The initial nanoparticles were constructed using the ASE suite~\cite{ase-paper} and were placed in the centre of the box. The NbP, NbAs and TaAs nanoparticles contained 9 unit cells in a 3x3x1 arrangement (72 atoms) along the x, y and z-direction, whereas the TaS$_2$ nanoparticle consisted of 18 unit cells in a 3x3x2 arrangement (54 atoms). The initial crystal structures for NbP, NbAs, TaAs and TaS$_2$ were taken from Refs.~\cite{NbPstructure, NbAsstructure, TaAsstructure} and \cite{TaS2structure}, respectively. All nanoparticles were subsequently solvated with 1100 water molecules using Packmol~\cite{packmol}. The so prepared systems were simulated for 10~ps using a discretized timestep of 0.5~fs by means of classical molecular dynamics (MD) in the canonical NVT ensemble at 300~K to relax the water molecules around the nanoparticles. The interatomic interactions were modelled using the CHARMM force field in conjunction with the flexible TIP3P water model~\cite{tip3pfl}. In these simulations runs the nanoparticles were given zero charge, whereas the the missing Lennard-Jones parameters for Nb, P, Ta and S were assigned according to the Universal Force Field (UFF)~\cite{UFF}. 
The resulting structure was then used as a starting point for our second-generation Car-Parrinello-based QM/MM MD simulations \cite{Karhan}. The QM region contained only the nanoparticle in a cubic 22~\AA\ long periodic supercell, whereas the water molecules were treated at the MM level. Even though including the solvation water layer in the QM zone would be desirable, this is complicated by the fact that the water molecules at finite-temperature are constantly transitioning between the QM and MM regions, which would necessitate the usage of sophisticated adaptive-resolution schemes~\cite{Praprotnik2005, Ensing2007, Bernstein2015, DelleSite2018}. The interactions between the MM and QM parts was calculated at the QM level, using the Gaussian expansion of the electrostatic potential (GEEP) method in conjunction with the electrostatic coupling of QM periodic images as developed by Laino et al. \cite{QMMM1,QMMM2}. To accelerate the computationally dominating ab-initio MD of the QM region, the second-generation Car-Parrinello MD scheme of K\"uhne et al. was employed \cite{CP2G,CP2Grev,Prodan2018}. Using the Gaussian and plane waves (GPW) approach \cite{CP2Krev}, the Kohn-Sham orbitals were expanded in contracted Gaussians, whereas the electronic charge density was represented using plane waves. The former was expanded in a molecularly optimized double-$\zeta$ basis with one additional set of polarization functions (DZVP) \cite{molopt}, while for the latter, a density cutoff of 240~Ry was used. The core electrons were represented by norm-conserving Goedecker-Teter-Hutter (GTH) pseudopotentials \cite{GTH1,GTH2,GTH3}, and  unknown exchange and correlation potential substituted by the PBE generalized gradient approximation~\cite{PBE}. With these settings, all systems were again equilibrated for 5~ps in the NVT ensemble, followed by a 50~ps long production run. For that purpose, the CSVR thermostat of Bussi et al. was separately applied to the MM and QM regions, with a time constant of 50~fs \cite{CSVR}.

\section{Results and discussion}
\label{results}
We begin the discussion of our results by examining the macroscopic properties of the studied nanocrystals, namely the solvent accessible surface area (SASA) and the volume. In order to compute the SASA and the volume, radical Voronoi tesselation, which takes atomic radii into account, was employed using the Voro++ library~\cite{voroplpl}. 
\begin{figure}
	\centering
	\includegraphics[width=\linewidth]{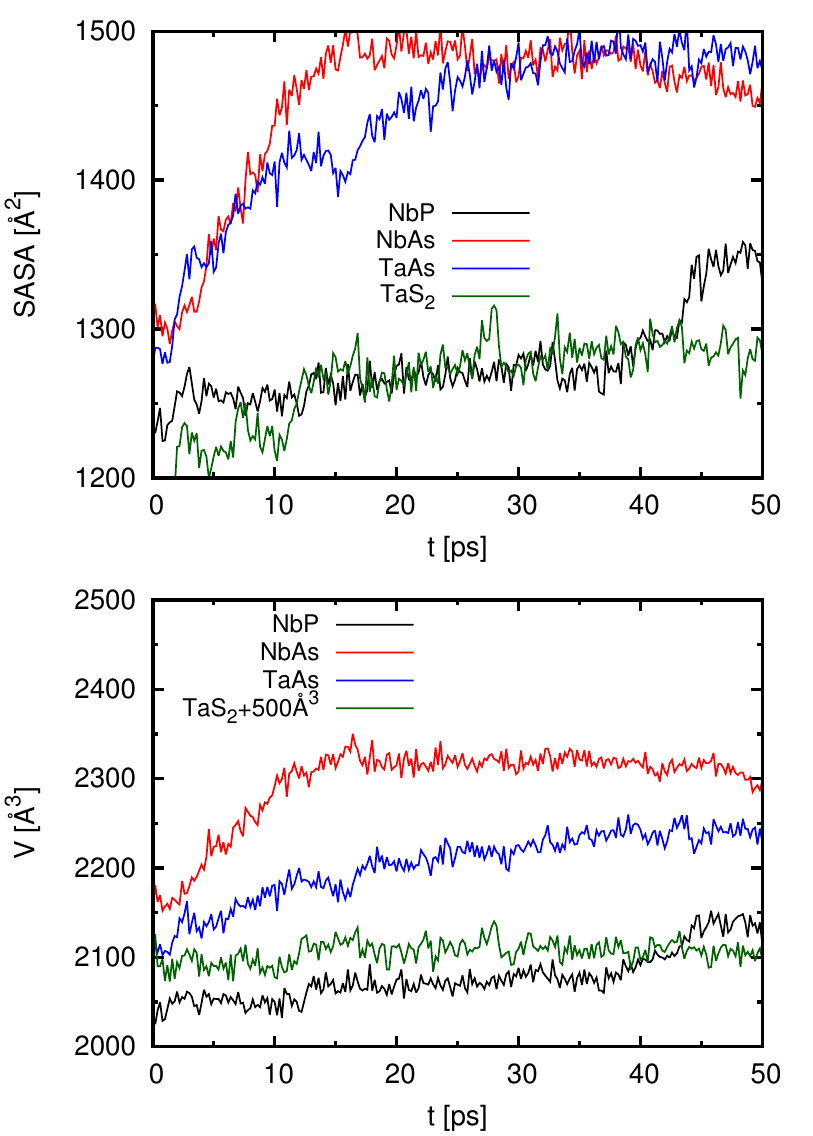}
	\caption{Solvent accessible surface area (top) and volume of the nanoparticle (bottom) as a function of time during the production run.}
	\label{fig:sasavol}
\end{figure}
The results are shown in Fig.~\ref{fig:sasavol} and reveal that the monopnictides nanoparticles we have considered here have larger volumes and also a higher SASA than TaS$_2$, which is in line with the larger nanoparticle size, as stated above. Furthermore, both the volume and SASA of the monopnictides decrease in the order NbAs $>$ TaAs $>$ NbP, which is exactly reverse to the order their activity in HER decreases.\\
Following the analysis of the macroscopic properties, the microscopic structure of the water molecules around the nanoparticles was investigated. To that extent, the last 35~ps of the production runs were used to collect statistics for all structure-related properties discussed hereafter. In Fig.~\ref{fig:grO} the partial radial distribution functions $g(r)$ (RDF) of metallic and nonmetallic atoms of the nanoparticles and water oxygen atoms are depicted, the values of which are related to the probability of finding a water molecule at a certain distance $r$ from a metallic and nonmetallic surface atom, respectively \cite{PCF}.
\begin{figure}
	\centering
	\includegraphics[width=\linewidth]{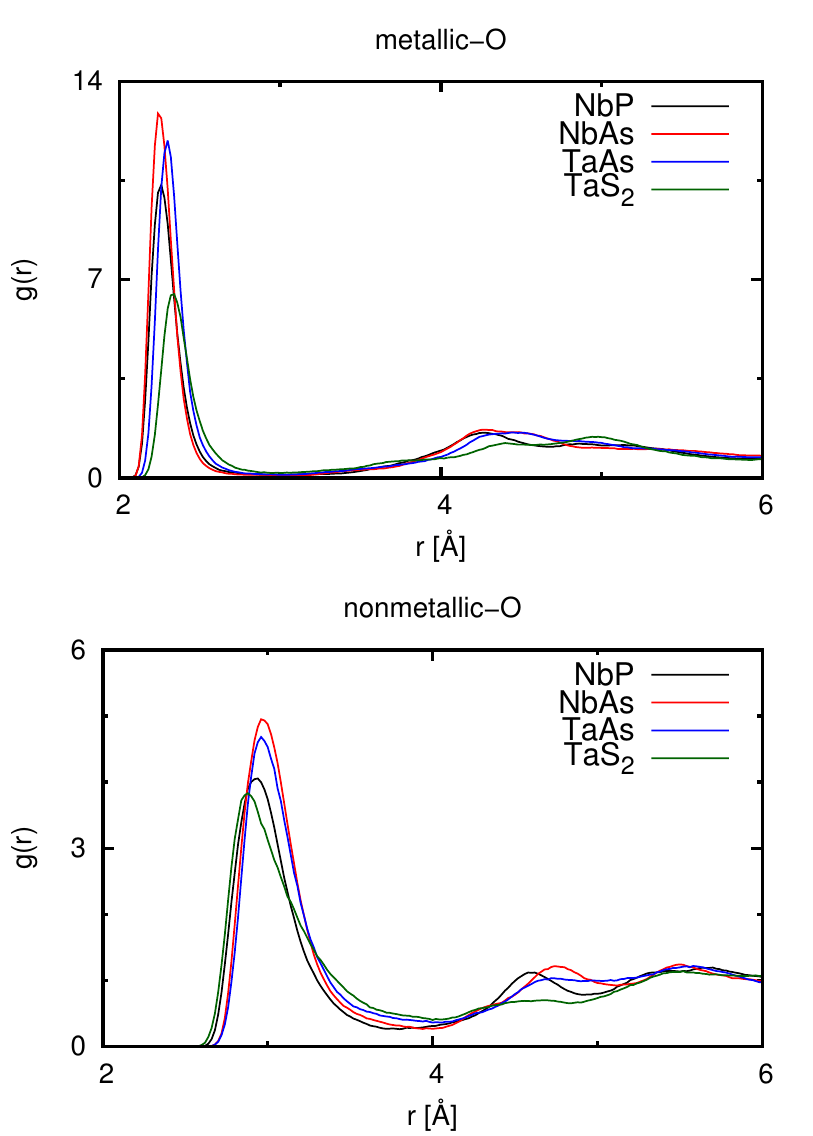}
	\caption{Radial distribution functions the oxygen atoms of water and metallic (top), as well as nonmetallic (bottam) atoms.}
	\label{fig:grO}
\end{figure}
For both classes of RDFs, the observed trends are identical. The intensity of the first peak decreases in the order NbAs $>$ TaAs $>$ NbP $>$ TaS$_2$. Looking at the monopnictides only, this accounts for the higher water affinity of NbAs in comparison to NbP. Since TaS$_2$ has a lower number of atoms compared to the other three systems we have investigated here, comparing its peak intensity with that of the monopnictides is meaningless. Instead, the corresponding integrals that represents the total coordination numbers must be taken into account. To be specific, the average number of water molecules $N_c$ around an atom inside a sphere with the radius $r_m$ can be obtained as $N_c=4\pi\rho_{\mathrm{O}}\int_0^{r_m} g_{\mathrm{O}}(r)r^2\,\mathrm{d}r$, where $\rho_{\mathrm{O}}$ is the average number density of oxygen atoms from water molecules and $g_{\mathrm{O}}$ the RDF of these oxygen atoms, which are at a distance $r$ from the corresponding surface atoms. The number of nearest neighbours is obtained by integration of the RDF plots up to the first minimum. Since not all nanoparticle atoms can be attributed to the surface, only those RDFs obtained for nanoparticle atoms having at least one oxygen atom within the first minimum distance (cf. Fig.~\ref{fig:grO}) were considered. The corresponding results for the investigated systems are summarized in Table~\ref{tab:coordnum}. 
\begin{widetext}
\begin{table}
    \centering
    \caption{Average number of coordinated water molecules $N_c$ around metallic (X) and nonmetallic (Y) atoms on the surface of the nanoparticles, as obtained by integration of the corresponding RDF up to the first minimum (the values of the minima are given in brackets). Note that the total is not the sum metallic and nonmetallic parts, as a single water molecule can be a neighbour of more than one nanoparticle atom.}
    \label{tab:coordnum}
    \begin{tabular}{c|c|c|c}
       nano- & $N_c$ of the nonmetallic part & $N_c$ of the metallic part  & total $N_c$ \\ 
             particle & [Minima Positions Y-O, \AA] & [Minima Positions X-O, \AA]& [Minima, \AA]\\\hline
     NbAs  & 5.25 [3.94] & 8.53 [3.04] & 6.89 [3.64] \\ 
     TaAs & 5.33 [3.94] & 8.47 [3.04] & 6.96 [3.68]\\
     NbP & 4.82 [3.80] & 7.20  [3.04]& 5.97 [3.54] \\
     TaS$_2$ & 3.83 [4.00] & 8.56 [3.04] & 7.75 [4.00]
    \end{tabular}
\end{table}
\end{widetext}
These data give evidence that the water coordination around the surface atoms of the NbP nanoparticle, as the most active HER catalyst, is significantly lower (total 5.97) than for the other two monopnictides ($\approx$7). Our results also suggest that the lower coordination number is mainly originating from the low coordination of the metallic part. By contrast, TaS$_2$, which exhibits the lowest performance in HER, has by far the highest number of water molecules coordinated to each surface atom, which again is due to the coordination of the metallic part. 

\begin{figure}
	\centering
	\includegraphics[width=\linewidth]{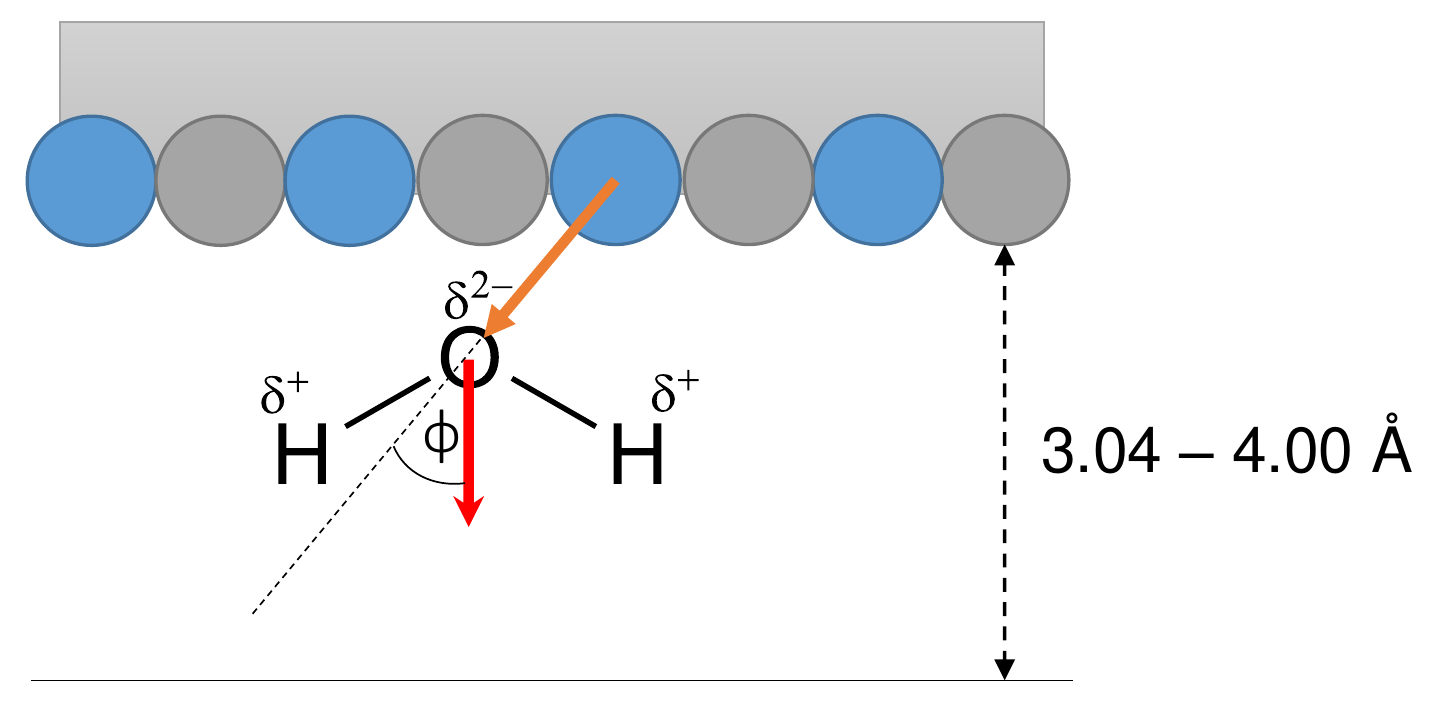}
	\caption{Illustration to define the angle $\phi$ between the water dipole vector and the vector starting at the centre of a surface atom and pointing towards the water oxygen atom.}
	\label{vector}
\end{figure}

\begin{figure}
    \centering
    \includegraphics[width=\linewidth]{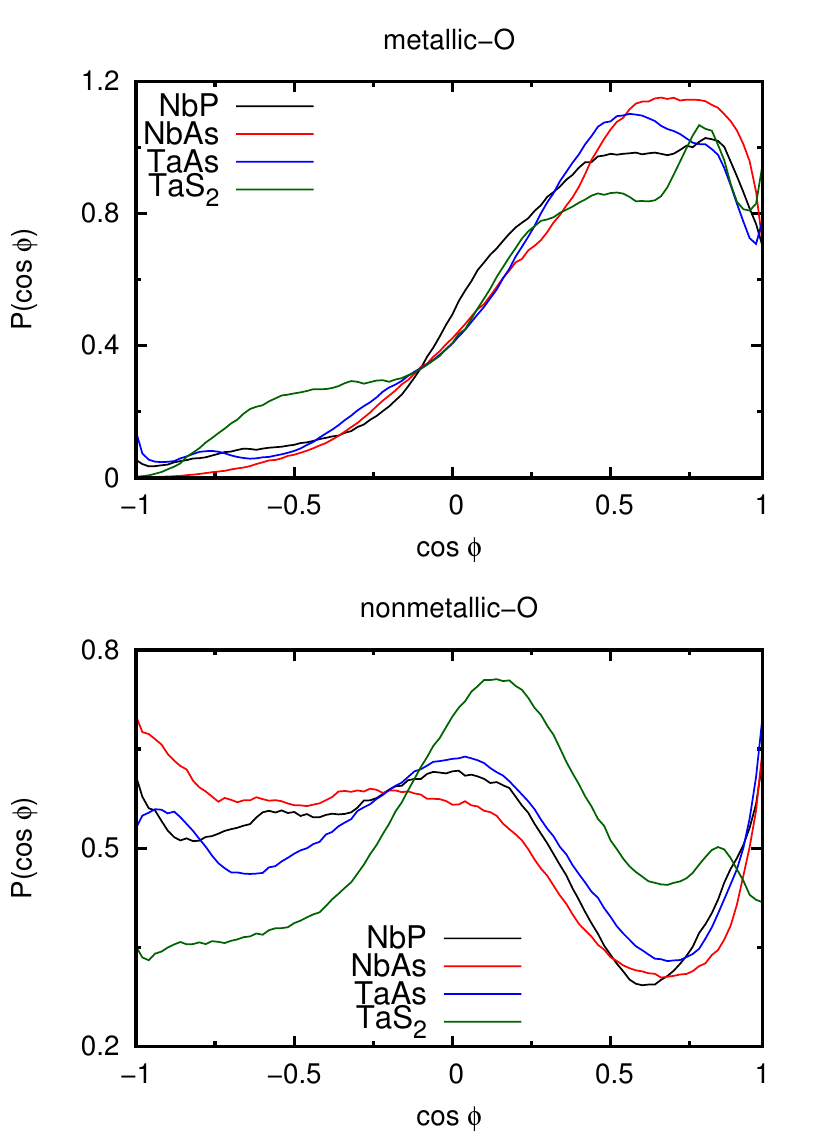}
    \caption{Distribution $P$ as a function of the cosine of the angle $\phi$ between the water dipole moment vector within the first minimum of the corresponding RDF and the vector starting at the metallic atom (top) or nonmetallic atom (bottom) and pointing towards the water oxygen atom. All integrals of the curves are normalized to unity.}
    \label{fig:distO}
\end{figure}

Besides the coordination numbers, the number of hydrogen bonds (HBs) per water molecule in contact with the surface of the nanoparticle was calculated. This involves all water molecules that are neighbored to at least one nanoparticle atom within the distance of 4~\AA. More precisely, we distinguish between HBs, which are only formed between water molecules in contact with the surface (surface-surface), and HBs that are formed between the surface-bound water molecules and the bulk (surface-bulk).
A simple geometric criterion was applied to determine the hydrogen-bonded water molecules: it was assumed that the distance between the oxygen atoms of donating and accepting water molecules is less than 3.5~\AA, and simultaneously the angle between the O-O axis and one of the O-H bonds is less then 30$^{\circ}$ \cite{Luzar1, Luzar2, CPwater}. The so obtained results are summarized in Table~\ref{tab:hbonds}.
\begin{table}
    \centering
    \caption{Average number of HBs between water molecules in contact with the surface (surface-surface) and the average number of hydrogen bonds between surface-bound water molecules and the bulk (surface-bulk).}
    \label{tab:hbonds}
    \begin{tabular}{c|c|c}
      nanoparticle & surface-surface & surface-bulk \\ \hline
     NbAs  &1.7$\pm$0.1 &2.4$\pm$0.1 \\ 
     TaAs & 1.85$\pm$0.08 & 2.52$\pm$0.08 \\
     NbP & 1.8$\pm$0.1 & 2.5$\pm$0.1\\
     TaS$_2$ & 2.24$\pm$0.08 & 2.96$\pm$0.09 
    \end{tabular}
\end{table}
We find that the monopnicitides we have considered exhibit a similar number of HBs near the surface, whereas the corresponding number of HBs per water molecule near the TaS$_2$ surface is increased by approximately 0.5 (cf. Table~\ref{tab:hbonds}). This observation is consistent with the high water affinity of TaS$_2$ as already alluded to above and has been correlated with its low catalytic performance. In other words, the higher the number of HBs the lower its activity. In addition to the number of HBs near the surface, the orientation of the water molecules around the bulk surface is analyzed in terms of the angle $\phi$ between the water dipole vector and the vector starting at the centre of a surface atom and pointing towards the water oxygen atom, as illustrated in Fig.~\ref{vector}. 
The distribution as a function of the cosine of the angle $\phi$ is plotted in Fig.~\ref{fig:distO}. 
Again, as before, only those water molecules that are closer than the first minimum of the corresponding RDF with the water oxygen atoms are considered. Our simulations yield rather similar orientations for the monopnictides, especially for the metallic atoms of all considered nanoparticles. They all have a preferential orientation between $\cos\phi\approx0.45$ and $\cos\phi\approx0.85$, which corresponds to $\phi$ being between $\sim 32^{\circ}$ and $\sim 63^{\circ}$. Nevertheless, contrary to NbAs and TaAs, 
NbP exhibits a slightly wider distribution, which corresponds to a less ordered water framework near the surface. The water molecules around the TaS$_2$ nanoparticles, however, have at least two preferential orientations, with the one at $\cos\phi\approx0.8$ ($\phi\approx 37^{\circ}$) being significantly more pronounced than the broad distribution at $\cos\approx -0.5$ ($\phi\approx 120^{\circ}$). 

Taking the nonmetallic nanoparticle atoms into account, the here considered monopnictides again obey a very similar behaviour. As can be seen in the bottom panel of Fig.~\ref{fig:distO}, the O-H bonds of water are mostly oriented towards the nonmetallic atoms, which corresponds to $\cos\phi < 0$. All three monopnictide systems have a pronounced minimum between $\cos\phi\approx 0.6$ ($\phi\approx 53^{\circ}$) for NbP and $\cos\phi\approx 0.7$ ($\phi\approx 45^{\circ}$) for NbAs and TaAs. Interestingly, also a high proportion of parallely oriented vectors is observed. However, integration of the corresponding distribution functions reveals that this corresponds only to roughly 2\% of the water molecules. By contrast, the angle distribution for TaS$_2$ is qualitatively different and shows a broad peak at around $\cos\phi\approx 0.15$ ($\phi\approx 81^{\circ}$) and a second less pronounced peak at $\cos\phi\approx 0.85$ ($\phi\approx 32^{\circ}$), which suggests the existence of two preferential orientations of the water molecules. 

However, we expect that gaining further insights regarding the impact of locality and strength of possible water-mediated hydrophilic interactions on the observed catalytic activity requires the combination novel methods based on the collective, long-wavelength electrostatic response of water to such surfaces~\cite{Remsing2015}, as well as condensed-phase energy decomposition analyses~\cite{ALMOrev, ALMOnatureI, ALMOnatureII}. 


\section{Experimental part}
\label{expdet}
\begin{figure}
    \centering
    \includegraphics[width=\linewidth]{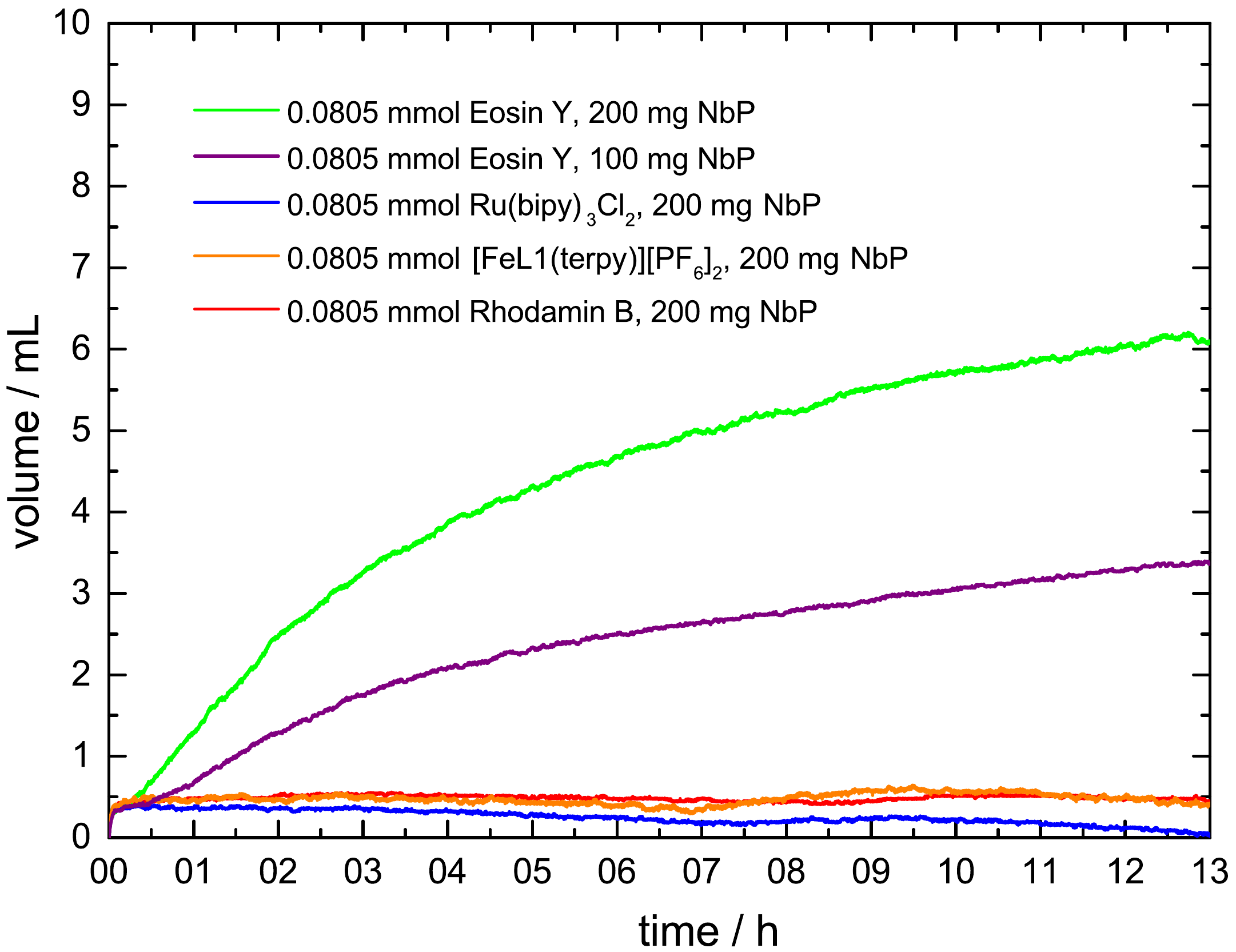}
    \caption{The volume of hydrogen gas as a function of time for different dyes used.}
    \label{fig:expplot}
\end{figure}

In addition to our aforementioned theoretical calculations, the influence of the dye used on the photocatalytic evolution of hydrogen gas was investigated experimentally. Based on our computational findings, the most effective HER catalyst NbP was employed and the produced gas volume measured using a previously descirbed experimental set-up~\cite{mb1}. In all of our experiments, an excess of NbP (1.61 mmol, 200 mg) was employed as photocatalyst with 81 $\mu$mol of dye. Both the dye and catalyst were dissolved in 19.2 mL of a triethanolamine (15 v/v) aqueous solution. In comparison to the original experiment, this corresponds to a halved concentration of the catalyst and a 14.4 times increased dye concentration~\cite{fpaper}. After 13 hours of irradiation, 1140 $\mu$mol$g^{-1}$ gas were detected (green curve in Fig.~\ref{fig:expplot}). The amount of hydrogen per gram of catalyst is similar to that of powdered single-nanoparticle NbP after 3 hours of irradiation reported by Rajamathi et al. \cite{fpaper}. The catalytic system investigated herein shows a lower activity because the plateau in hydrogen production is reached after a longer period of time. Due to the significantly increased dye loading, the increased catalytic performance is plausible and can be explained by the lower turnover frequency at the catalyst, which seems to be rate limiting in this catalyst cycle.    
To verify this hypothesis, the catalyst amount was again reduced by a factor of two while keeping the remaining parameters constant. This resulted in an equivalent decrease in the volume of evolved hydrogen gas (violet curve). Therefore, the number of active HER catalyst sites is obviously limiting the catalytic performance. In this context it is noteworthy that the photocatalytic system reaches its full catalytic activity after around 30 minutes induction time. This may be attributed to the water clusters bound to the surface, as discussed above.\\
Interestingly, with Rhodamin B as an alternative dye to Eosin Y, no catalytic activity was observed. According to available literature ~\cite{exper1}, this dye should exhibit activity with better long term stability of the reaction. In addition to organic dyes, metalorganic photosensitizers were investigated. A prominent standard example is [Ru(bipy)$_3$]Cl$_2$. Just as Rhodamin B, this photosensitizer turned out to be inactive (cf. Fig.~\ref{fig:expplot}, blue curve). The addition of methyl viologene, which acts as an efficient redox mediator, also did not result in any catalytic activity. In the same way, an iron-based system [FeL1(terpy)][PF$_6$]$_2$ 
with L1 = 2,6-bis[3-(2,6-diisopropylphenyl)imidazol-2-ylidene]pyrazine and terpy = 2,2':6':2''-terpyridine was found to be inactive, although the pyrazine nitrogen can allow a more efficient interaction with the NbP catalyst. It is commonly accepted that the catalytic activity is correlated to the lifetime of the photoactive state defines. Accordingly, the photosensitizer has to be quenched by the catalyst or triethanolamine to allow an electron transfer in a diffusion controlled process. This consideration obviously does not apply here. Although the excited state lifetime of Eosin Y (triplet state, 24 $\mu$s) \cite{dye2} and [Ru(bipy)$_3$]Cl$_2$ (MLCT state, 1~$\mu$s)~\cite{exper4} are of the same order of magnitude, their catalytic activity is completely different. Similarly, although Rhodamin B is characterized by a comparatively short lifetime in the low ns range \cite{dye1}, its photocatalytic behaviour should still be different from the iron-based photosensitizer with a small ps-lifetime \cite{Zimmer}.\\
If we consider the binary behaviour of the employed photosensitizers with respect to the proton reduction activity, a particular discriminating order-parameter appears to be more appropriate to understand the observations. The most striking difference between Eosin Y and all other photosensitizers is that the former one carries a negative charge, while the later ones are all positively charged. 
Although other order-parameters like redoxpotentials (including those of the excited states) have to be further considered as well, the interaction of the photosensitizers with the Weyl-Semimetal including electrostatic interactions need further in-depth elucidation. In particular, the pH value is important, since it influences not only dyes and sacrificial reductants, but also the effects of water molecules around the surface as discussed above.

\section{Conclusions}
\label{conclusions}
In conclusion, for the first time the catalytic activity of Weyl-semimetals has been investigated by employing QM/MM MD simulations. Considering an identical number of nanoparticle atoms, NbAs and TaAs are characterized by a larger SASA and volume of the nanoparticle than NbP. Therefore, the higher catalytic activity of NbP can be correlated with its larger specific surface.
In addition, the surface atoms of NbAs, TaAs and TaS$_2$ are coordinated by more water molecules than the surface atoms of NbP. Consequently, a large water cluster bound to the surface results in a diminished catalytic activity. Furthermore, the number of hydrogen bonds in the layer adjacent to the surface was studied. While the number of hydrogen bonds for monopnictides is nearly identical, TaS$_2$ exhibits a higher hydrogen bond count. In agreement with this statement, water molecules around NbP seem to have slightly less pronounced preferential orientations than in the case of the other monopnictides. In contrast, water molecules around TaS$_2$ have the most pronounced preferential orientations among the four materials investigated herein. Therefore, a lower ordered water cluster at the surface results in an increased HER activity.\\
In order to complement these theoretical results, we investigated the influence of different dyes on the catalytic performance of NbP as photocatalyst in catalytic experiments. As an outcome, the concentration of NbP and the number of active sites, which are formed after a certain induction time, are rate limiting. More importantly, the chemical nature of the dye itself is crucial for the volume of hydrogen produced. Anionic Eosin Y turned out to be the only active one. By contrast, cationic dyes are inactive, no matter if they are of metalorganic (e.g. [Ru(bipy)$_3$]Cl$_2$) or organic (e.g. Rhodamin B) in nature. Most surprisingly, the excited state lifetime shows no correlation with the photocatalytic proton reduction activity. Future research now focuses on a combined theoretical and experimental approach in order to investigate the electrostatic factors influencing the interaction of dyes with the surface atoms and the proton reduction process at the surface itself.

\begin{acknowledgments}
This project has received funding from the European Research Council (ERC) under the European Union's Horizon 2020 research and innovation programme (grant agreement No 716142) and by the Federal Ministry of Education and Research of Germany (BMBF) in the framework of the german-swedish R{\"o}ntgen-{\AA}ngstr{\o}m-Cluster `SynXAS' (FKZ: 05K18PPA).
The generous allocation of computing time on the FPGA-based supercomputer ``Noctua'' by the Paderborn Center for Parallel Computing (PC$^2$) is kindly acknowledged.
\end{acknowledgments}

%

\end{document}